\documentclass[twocolumn, twocolappendix]{aastex631}%linenumbers,

\usepackage{multirow}
\usepackage{booktabs}
\begin{document}

\title{Cosmology from Point Clouds with Dark Matter Halos from the Quijote Simulations}

\correspondingauthor{Atrideb Chatterjee}
\email{atrideb.chatterjee@iucaa.in}

\author[0000-0002-0159-8708]{Atrideb Chatterjee}
\affiliation{Inter-University Centre for Astronomy and Astrophysics, Post Bag 4, Ganeshkhind, Pune 411007, India}
\affiliation{Kapteyn Astronomical Institute, University of Groningen, PO Box 800, 9700 AV Groningen, The Netherlands}

\author[0000-0002-4816-0455]{Francisco Villaescusa-Navarro}
\affiliation{Center for Computational Astrophysics, Flatiron Institute, 162 5th Avenue, 
New York, NY, 10010, USA}
\affiliation{Department of Astrophysical Sciences, Princeton University, 4 Ivy Lane, Princeton, 
NJ 08544 USA}

%% AASTeX 6.31 has the new \collaboration and \nocollaboration commands to

%% Mark off the abstract in the ``abstract'' environment. 
\begin{abstract}

We train a novel deep learning architecture to perform likelihood-free inference on the value of the cosmological parameters from halo catalogs of the \texttt{Quijote} N-body simulations. Our model takes as input a halo catalog where each halo is characterized by its position, mass, and velocity modulus. By construction, our model is E(3) invariant and is designed to extract information hierarchically. Unlike graph neural networks, it does not require the transformation of the input halo (or galaxy) catalog into a graph. Given its simplicity, our model can process point clouds with large numbers of points. We discuss the advantages of this class of methods but also point out its limitations and potential ways to improve them for cosmological data.

\end{abstract}

 
%%%%%%%%%%%%%%%%%%%%%%%%%%%%%%%%%%%%%%%%%
\section{Introduction} 
\label{sec:intro}
Our ability to characterize and understand our Universe depends on the level of accuracy to which the value of the different cosmological parameters can be constrained. The spatial distribution of galaxies in the sky is influenced by various cosmological parameters. As a result, galaxy clustering has become a common practice to study the laws and components of the Universe \citep{Uhlemann_2020, Villaescusa_Navarro_2020, Gualdi_2021, Valgiannis_2021, Liu_2022, Ajani_2020}. Upcoming sky surveys—such as CMB-S4 \citep{2022arXiv220308024A}, Roman Space Telescope \citep{2023arXiv230612514L}, Dark Energy Spectroscopic Instrument (DESI) \citep{DESI}, Euclid \citep{Laureijs2011, Euclid2016, Euclid2022-Tiago_Castro}, the Large Synoptic Survey Telescope (LSST) \citep{2019ApJ...873..111I}, J--PAS \citep{Benitez2014} will gather data with unprecedented precision. This will make galaxy clustering an especially powerful method for advancing precision cosmology.

In view of this, a large number of efforts using different machine learning techniques \citep{Siamak_16, Fluri_19, 2022OJAp....5E..18M, Niall_2020, 2019MNRAS.484..282G, Hortua_2021, 2022ApJ...937..115V, GNN_MW_M31, 2022OJAp....5E..18M, Hasan_HIFLOW, 2022arXiv221112346A, Cosmology_with_One_Galaxy_Villaescusa-Navarro:2022twv,2023arXiv231117141C, 2023arXiv231117141C, Domain_adaptive_roncoli, 2024arXiv240205137H} have been put forward to make optimal use of these upcoming data. The unique ability of these techniques to learn extremely complicated features from a given dataset has given them a decisive advantage over traditional methods that employ summary statistics \citep{Marques_2019, Villaescusa_Navarro_2020, Gualdi_2020, Hahn_2020, Friedrich_2020, Giri_2020, Harnois_2021a, Samushia_2021,  2021MNRAS.500.5479B, Naidoo_2021, Bayer_2021, 2022arXiv220407646E}. On top of that, these methods can potentially address complex problems, like marginalizing over baryonic effects at the field level, that are quite challenging with standard methods \citep{2022ApJ...937..115V, Natali_2023}.

Data collected from telescopes are often released as catalogs of sources containing their positions and different properties rather than images of a particular area of the sky. These galaxy catalogs, with their inherent sparse and irregular nature, are difficult/suboptimal to use as the input for any image-based network\footnote{On one hand, one can place the observed catalog in a regular 2D or 3D grid at the expense of throwing away information on scales below the grid size. On the other hand, if the grid is sufficiently fine (i.e., containing at most one point per pixel), it will be a big grid filled out with zeros, given the sparse nature of the data.}. To address this issue, a new method for sparse data, called point cloud analysis, has emerged as a new avenue for 3D data analysis \citep{2016arXiv161200593Q, 2018arXiv181204244S, 2020arXiv200401803X, 2022OJAp....5E..18M, GNN_MW_M31, Hordan_WL, Hordan_Euclidean_Equivariant, Rose_GNN}. Further, with its rapid improvement, point cloud analysis has become very popular, performing better or at par with other state-of-the-art image-based analysis methods \citep{7353481, 2017arXiv170602413Q, 9577763}. 

In this work, we employ a novel architecture that takes a point cloud as input and processes it efficiently using Multi-Layer Perceptrons (MLPs), whereas Graph Neural Networks (GNNs) rely on graph-based structures and message passing. A key advantage of this model is its ability to extract information hierarchically, enabling the efficient processing of relatively large point clouds. We note that some previous works exploited complicated geometrical feature extractors to process point clouds \citep{2019arXiv191006849L, 2020arXiv201209688G}. Recently, \cite{2022arXiv220207123M} proposed a novel method that takes point clouds as input and processes them in a very similar fashion as GNNs (i.e., with simple, flexible, and learnable geometrical kernels) without having to predefine a graph from the point cloud. They have shown that their method either outperforms or performs at par with other state-of-the-art point cloud networks. Further, due to employing a simple feature extractor, this network is lightweight and fast compared to other 3D data analysis methods.

In this work, we perform parameter inference from point clouds using the method proposed by \cite{2022arXiv220207123M} but enhancing it to make it E(3) invariant. Our dataset is composed of halo catalogs from N-body simulations of the \texttt{Quijote} suite \citep{2020ApJS..250....2V}. We show that this network performs well when constraining the value of some parameters but fails on others. Besides, its performance against classical summary statistics is not good. We discuss in detail the limitations of the model and potential avenues to improve it.

The outline of the paper is as follows. We describe the data we use in Section \ref{sec:data}. In Section \ref{sec:methods}, we describe the model architecture, the way we impose the symmetries, the used loss function, the training procedure, and the validation metrics. We show the results we achieve from training the networks in Section \ref{sec:result}. We then discuss the limitations and future directions in Section \ref{sec:discussion}. We draw the main conclusions of this work in Section \ref{sec:conclusions}.

\section{Data} 
\label{sec:data}

In this paper, we work with 3D point clouds. A point cloud contains N points, and each point is characterized by its 3D position (x,y,z). The points can also have other properties, such as masses and peculiar velocity moduli \footnote{This is defined as $v = \sqrt{v_{x}^2+v^2_{y}+v^2_{z}}$ where $v_{x}$, $v_{y}$, $v_{z}$ are the cartesian components of the peculiar velocity respectively}. Each point cloud is fully described by five labels: the values of the cosmological parameters. In this work, we will, however, restrict ourselves to just two of these parameters: $\Omega_{\rm m}$ and $\sigma_8$.

A point cloud is constructed as follows. First, all halos from an N-body simulation are read. Next, we take the $N$ more massive halos, and for each halo, we store its position, mass, and velocity moduli. The labels of the point cloud are the value of the cosmological parameters of that halo catalog.

The halo catalogs are obtained from the \texttt{Quijote} simulations \citep{2020ApJS..250....2V} at $z=0$. In particular, we use the fiducial-resolution\footnote{Using the higher-resolution latin-hypercube will not have an impact on the results of this paper as we only use the most massive halos that are well resolved at this resolution.} latin-hypercubes that contain 2,000 N-body simulations whose parameters are varied within:
\begin{eqnarray}
0.1 \leq &\Omega_{\rm m}& \leq 0.5\\
0.03 \leq &\Omega_{\rm b}& \leq 0.07\\
0.5 \leq &h& \leq 0.9\\
0.8 \leq &n_s& \leq 1.2\\
0.6 \leq &\sigma_8& \leq 1.0
\end{eqnarray}

Thus, our dataset contains 2,000 point clouds, and each point cloud can have $N=1,024$, 4,096, or 8,192 points.

\section{Methodology} 
\label{sec:methods}

In this section, we describe our model's architecture, how we make it E(3) invariant, the loss function we employ, the training and validating procedure, and the validation metrics we use to quantify the model's performance.

\subsection{Architecture}
\label{subsec:architecture}

This work aims to infer the value of the cosmological parameters from the spatial distribution of dark matter halos of N-body simulations, represented as point clouds.

For each simulation, we have a collection of $N$ points $\{p_1, p_2,...p_n\}$ where each point represents a dark matter halo. We will denote the properties of the point $i$ by $f_i\in \mathbb{R}^d$. The properties of the $j$-th closest neighbor of the point $i$ are instead represented by $f_{i,j}\in \mathbb{R}^d$. The dimensionality of the input feature vector is 
\begin{itemize}
    \item $d=3$ when only considering halo positions,
    \item $d=4$ when using positions and masses,
    \item $d=5$ when using halo positions, masses, and velocity moduli.
\end{itemize}

Our model takes a point cloud as input and returns two values for each cosmological parameter $i$, $\mu_i$, and $\sigma_i$, representing the marginal posterior mean and standard deviation. 
\begin{equation}
g: \mathbb{R}^{N \times d} \to \mathbb{R}^{2\times N_\theta}~, 
\end{equation}
where $N_\theta$ is the number of cosmological parameters. The architecture of our models closely follows\footnote{While the model architecture is the same as in \cite{2022arXiv220207123M}, the loss function and the E(3)-invariant point features are modifications introduced in this work.} the \texttt{PointMLP-elite}\footnote{\url{https://github.com/ma-xu/pointMLP-pytorch}} model recently introduced in \cite{2022arXiv220207123M}, and consists of 1) an initial embedding, 2) a series of four \textit{stages}, and 3) an inference module. We show a scheme of the model architecture in Fig. \ref{fig:PointMLPElite}. We now describe each of these elements:

\begin{figure*}
\includegraphics[width =\textwidth]{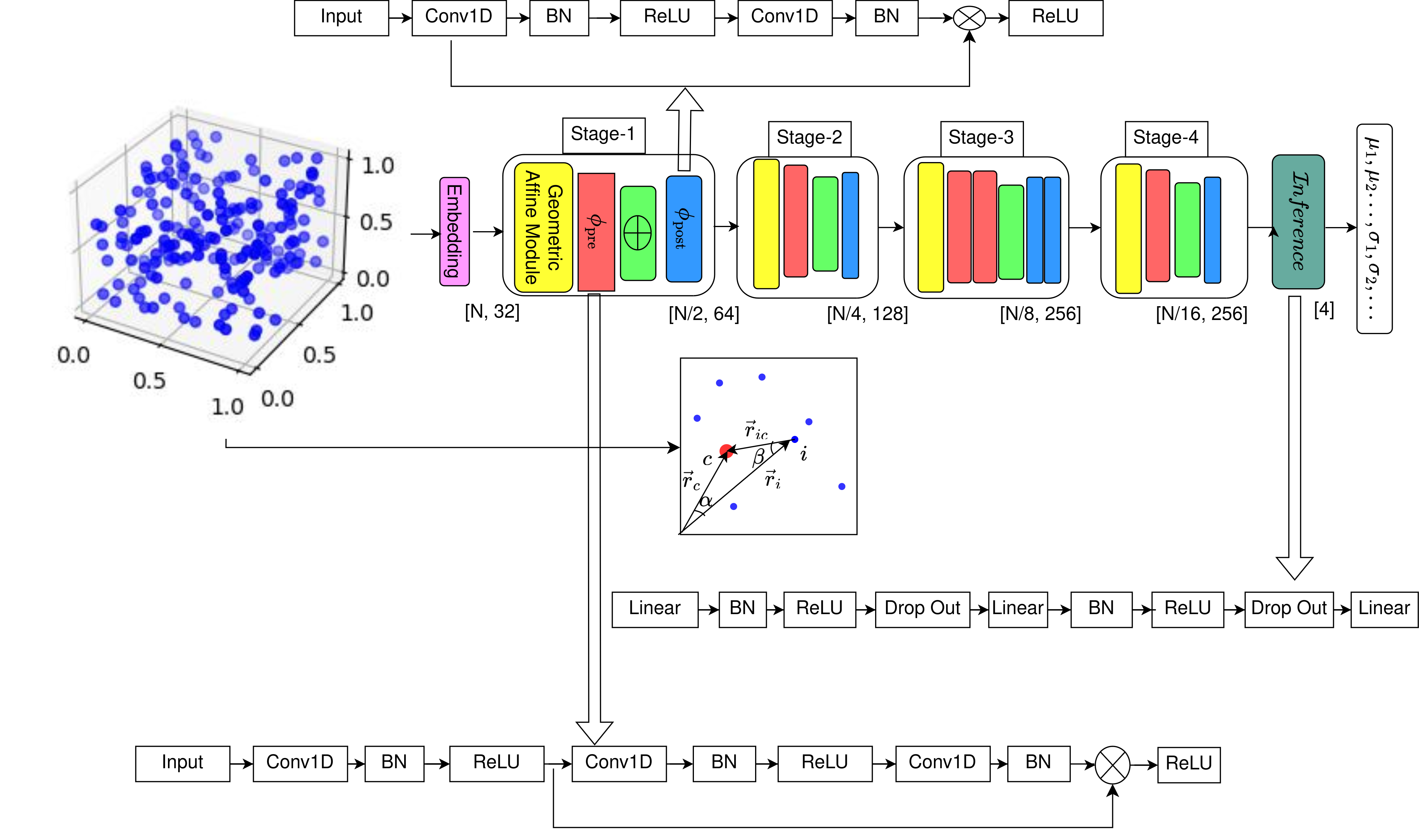}
    \caption{Structure of the network we use in this work. The model takes a point cloud consisting of $N$ points, each of them characterized by 3D positions (and masses and velocity moduli). The 3D position of each point is expressed as a distance and 2 angles (see central panel), making our model E(3) invariant. The embedding layer performs a 1D convolution and transforms the dimensionality of every point to 32. Next, the point cloud is passed through four \textit{stages} that capture the geometrical information on different scales. Finally, an inference layer consisting of MLPs will transform the data into a vector containing the posterior mean and standard deviation of each considered parameter.}
    \label{fig:PointMLPElite}
\end{figure*}

\begin{itemize}
\item \textbf{Embedding}. A 1D convolution that is applied to every point in the catalog and whose purpose is to embed the catalog in a higher dimension (32 in our case). 

\item \textbf{Stage}. At the beginning of a stage, a subset of $n$ points is selected using the Farthest Point Sampling (FPS) algorithm. In our case, we select half of the input points. Next, the features of the selected point are updated as follows. 

Given a point $i$ at stage $l$ with features $f^l_i\in 
 \mathbb{R}^{m}$, the features of its $k$ nearest neighbors $f^l_{i,j}$ are first transformed using a 
\textit{geometric affine module}:
\begin{equation}
\hat{f}^l_{i, j} = \alpha \odot \frac{f^l_{i, j}-f_i^l}{\sigma +\delta} + \beta
\label{Eq:affine}
\end{equation}
where $\alpha, \beta\in\mathbb{R}^m$ are learnable parameters, $\odot$ is the Hadamard product and $\sigma$ is a normalization defined as
\begin{equation}
\sigma=\sqrt{\frac{1}{k\times n\times m} \sum_{i=1}^n\sum_{j=1}^k (f_{i,j}^l-f_i^l)^2}~.
\end{equation}
$\delta=10^{-5}$ is added for numerical stability. As mentioned in \cite{2022arXiv220207123M} the affine transformation is introduced for transforming the local points to a normal distribution which in turn makes the network more robust, stable and accurate. Next, we update the features of point $i$ using the transformed properties of its $k$ nearest neighbors via 
\begin{equation}
f_i^{l+1}=\Phi_{\rm pos} \bigoplus_{j\in \mathcal{N}_i}\Phi_{\rm pre}(\hat{f}^l_{i, j})~,
\end{equation}
where $\Phi_{\rm pos}$ and $\Phi_{\rm pre}$ are the so-called PreBlock and PostBlock modules. Those contain 1D convolutions, batch norm, activation, and residual layers (see Fig. \ref{fig:PointMLPElite} for details). $\bigoplus$ is a permutation invariant operation (the maximum in our case) that operates over the neighbors of point $i$, $\mathcal{N}_i$. After one stage, we increase the dimensionality of $f_i^{l+1}$ by a factor of 2 with respect to $f_i^l$.

The features of the points are updated through 4 different stages as illustrated in Fig. \ref{fig:PointMLPElite}. Note that in the third stage, we use two PreBlocks and two PostBlocks instead of one. We emphasize that due to the hierarchical structure of the network if we start with $N$ points, we will have only $N/2^4$ points available after the four stages. We also note that the number of points to subsample at each stage can vary and does not need to be set to one-half.  

\item \textbf{Inference}. This part contains several MLP layers (Linear + batch norms + activations + dropout) (see Fig. \ref{fig:PointMLPElite}) that will transform the remaining $N/2^4$ points from the last stage into $2\times N_\theta$ numbers representing the first two moments of the posterior for each parameter.

\end{itemize}

\subsection{Symmetries}

The values of the cosmological parameters of a set of dark matter halos are not affected if we translate and/or rotate their positions. These symmetries can be learned in our model through data augmentation or imposed directly into the model. In our case, we decide to impose them using invariant input features of the point cloud. We checked that the data augmentation (random rotation, translation etc) increases the training time of the model significantly, where as making the input features E(3) invariant  bypasses the need for any data augmentation and makes the training much faster.

To achieve this, we take inspiration from \cite{GNN_CAMELS} and we transform the cartesian position coordinate of each halo into a new E(3) invariant coordinate system (see Fig. \ref{fig:PointMLPElite}) following these steps:
\begin{enumerate}
    \item We first calculate the position of the centroid of the distribution of the halos\footnote{We note that in principle we can take any other point inside the box.} (red point denoted by c in the bottom panel of Fig \ref{fig:PointMLPElite}). Next, we define a vector $\Vec{r}_{ic}$ joining the i-th halo to the centroid, i.e., $\Vec{r}_{ic} = \Vec{r}_{i} - \Vec{r}_{c}$, where $\Vec{r}_{i}$ and $\Vec{r}_{c}$ are respectively the positions of i-th halo and the centroid of the point cloud system.
    \item  Next, the distance $d_{ic} = |\Vec{r}_{ic}|$ between the centroid and the i-th halo is computed.
    \item We then determine the cosine of the angle $\alpha$ and $\beta$ (as shown in the bottom panel of Fig. \ref{fig:PointMLPElite}) by taking the scalar product between the vectors $(\Vec{r}_{i}$, $\Vec{r}_{c})$ and $(\Vec{r}_{i}, \Vec{r}_{ic})$ respectively, i.e.,
    $\cos{\alpha} = \Vec{r}_{i} \cdot \Vec{r}_{c}/|\Vec{r}_{i}||\Vec{r}_{c}|$  and $\cos{\beta} = \Vec{r}_{i} \cdot \Vec{r}_{ic}/|\Vec{r}_{i}||\Vec{r}_{ic}|$.
    \item Finally, we represent the position of the i-th halo by $( d_{ic},\, \cos{\alpha}, \, \cos{\beta})$ and employ these coordinates as the features for the 3D position rather than the Cartesian coordinate $(x, y, z)$.

    \item We emphasize that we take into account periodic boundary conditions when computing distances and angles between vectors.
\end{enumerate}

One may wonder if the affine transformation of Eq. \ref{Eq:affine} may not make sense in the new coordinate system, given the fact that point properties will not be zero-normalized anymore. We have checked that our results are pretty insensitive to this.

%%%%%%%%%%%%%%%%%%%%%%%%%%
\subsection{Loss Function}  

Our model is trained to perform likelihood-free inference on the value of the cosmological parameters. Given a set of $N$ points, the above architecture outputs $2\times N_\theta$ numbers. For each parameter, the model predicts the marginal posterior mean ($\mu_i$) and standard deviation ($\sigma_i$), defined as:

\begin{eqnarray}
\mu_i(\mathcal{P}) &=& \int_{\theta_i} p(\theta_i | \mathcal{P}) \theta_i d\theta_i~,\\
&& \nonumber \\
\sigma_i^2(\mathcal{P}) &=& \int_{\theta_i} p(\theta_i | \mathcal{P}) (\theta_i - \mu_i)^2 d\theta_i~,
\label{Eq:mean_std}
\end{eqnarray}
where $\mathcal{P}$ represents a point cloud. This is accomplished by minimizing the following loss function \citep{moment_networks, CMD} 

\begin{eqnarray}\label{eqn:loss_function}
\mathcal{L}&=& \sum_{i=1}^{N_{\theta}}\log\left(\sum_{j\in{\rm batch}}\left(\theta_{i,j} - \mu_{i,j}\right)^2\right)\nonumber \\
+&& \sum_{i=1}^{N_{\theta}} \log\left(\sum_{j\in{\rm batch}}\left(\left(\theta_{i, j} - \mu_{i,j}\right)^2 - \sigma_{i,j}^2 \right)^2\right)~,
\label{Eq:loss}
\end{eqnarray}
where $\theta_{i,j}$, $\mu_{i,j}$, $\sigma_{i,j}$ represent the true, inferred mean, and inferred standard deviation of the parameter $i$ for the sample $j$. The reason for using this loss function is that it bypasses the need to compute the posterior density of each parameter, directly obtaining the estimates of moments of the marginalized distribution of all the parameters.

\subsection{Training procedure}

We split the 2,000 point clouds available into training (80\%), validation (10\%), and testing (10\%). Each point cloud can have $N=\{1024, 4096, 8192\}$ points and the number of neighbors is set to either $k = \{32, 64\}$. The features of the points can be $(i)$ their positions $f_i\in\mathbb{R}^3$ (referred to as Pos-only hereafter), $(ii)$ their positions and masses $f_i\in\mathbb{R}^4$ (hereafter Pos+M), or $(iii)$ their positions, masses, and velocity moduli $f_i\in\mathbb{R}^5$ (Pos+M+V hereafter). All features are normalized to have their values between 0 and 1. In the case of the masses, we first compute its $\log_{10}$, and then normalize them with min-max. We also normalize the value of the output features representing the cosmological parameters to be within 0 and 1.

Next, the point clouds in the training set are passed through the architecture depicted in Sec. \ref{subsec:architecture}, and the loss function of Eq. \ref{eqn:loss_function} is minimized through gradient descent. The network weights are optimized using the SGD optimizer \citep{2016arXiv160904747R} with a Nesterov momentum of 0.9. We also used the cosine annealing scheduler \citep{2016arXiv160803983L} that modulates the learning rate during training. We use a batch size of 32 and train the model for 250 epochs. 

We consider three hyperparameters for our model $(i)$ the maximum learning rate, ($ii$) the minimum learning rate, and ($iii$) the value of the weight decay. We optimize the value of these hyperparameters for each configuration (i.e., for a given total number of points and neighbors) using the hyperparameter optimization software \texttt{OPTUNA} \citep{2019arXiv190710902A}. The optimization is performed by minimizing the validation loss, and we carry out at least 100 trials. 

We have studied the impact of the total number of halos considered, $N$, and the number of closest neighbors, $k$, on our results. From now on, we will refer to a configuration containing $N$ halos with $k$ neighbors as $N_k$. For example, $1024_{64}$ denotes point clouds containing 1,024 halos and considering 64 neighbors.

%%%%%%%%%%%%%%%%%%%%%%%%%
\subsection{Validation metrics}

We employ four statistics to quantify the accuracy and precision of our models and compare our results with other recent studies performed by different groups. Let us consider a point cloud from the test set $\mathcal{P}_i$, which contains $N$ point clouds, each of them characterized by the true value of the $j$ cosmological parameters $\theta_{ij}$. When $\mathcal{P}_i$ is passed through the model, we will obtain a posterior mean and standard deviation for each parameter $\mu_{ij}$ and $\sigma_{ij}$.

We can use the below four metrics to quantify the accuracy and precision of the model:

\begin{itemize}
\item \textbf{The mean relative error, $\epsilon_{i}$} defined as 
\begin{equation}
    \epsilon_{i} = \frac{1}{N} \sum_{j = 1}^{N} \frac{|\theta_{ij} - \mu_{ij}|}{\theta_{ij}},
\end{equation}
The smaller the value of $\epsilon_i$, the more precise the network is.

\item \textbf{The coefficient of determination, $R_{i}^2$}, defined as
\begin{equation}
    R_{i}^2 = 1 - \frac{\sum_{j = 1}^{N} (\theta_{ij} - \mu_{ij})^2}{\sum_{j = 1}^{N} (\theta_{ij} - \overline{\theta}_{i})^2},
\end{equation}
where $\overline{\theta}_i=\sum_{j= 1}^{N} \theta_{ij}$. The closer the value of $R_i^2$ is to 1, the more accurate the network is.

\item \textbf{The mean squared error} MSE$_i$
is defined as

\begin{equation}
    {\rm MSE}_{i} = \frac{1}{N}\sum_{j = 1}^{N} (\theta_{ij} - \mu_{ij})^2
\end{equation}
The smaller the value of the mean squared error, the more accurate the network is.

\item \textbf{The $\chi^2$ value}, 
 $\chi_{i}^2$ that is defined as 
\begin{equation}
\chi_{i}^2=\frac{1}{N}\sum_{j = 1}^{N} \frac{(\theta_{ij} - \mu_{ij})^2}{\sigma_{ij}^2}
\label{Eq:chi2}
\end{equation}
We use this statistic to quantify the quality of the network errors. Values close to 1 indicate that network errors are correctly calibrated.

\end{itemize}

%%%%%%%%%%%%%%%%%%%%%%%%%%%%%%%%%%%%%
%%%%%%%%%%%%%%%%%%%%%%%%%%%%%%%%%%%%%
\section{Results}
\label{sec:result}
\begin{figure*}
\begin{center}
\includegraphics[width=0.45\linewidth]{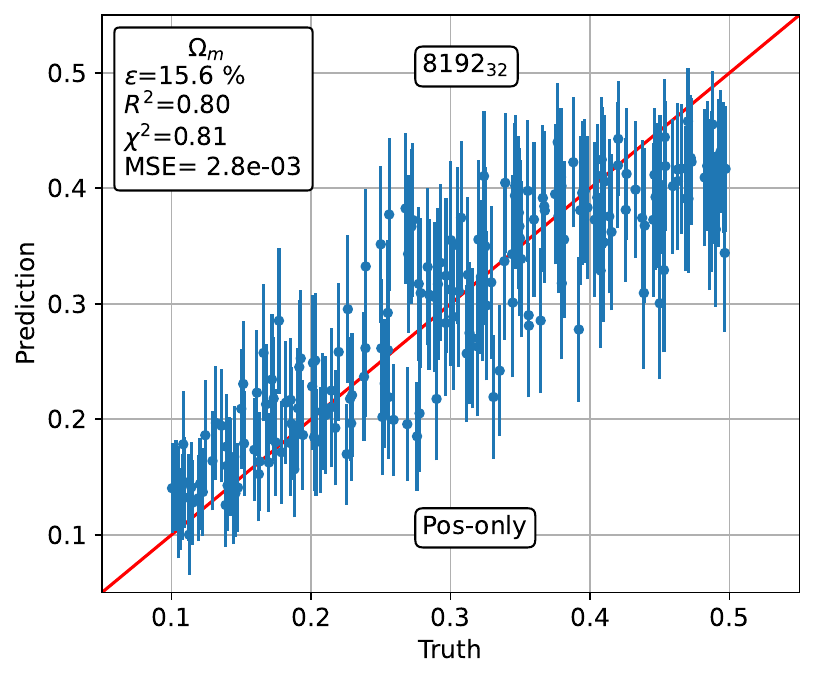}
\includegraphics[width=0.45\linewidth]{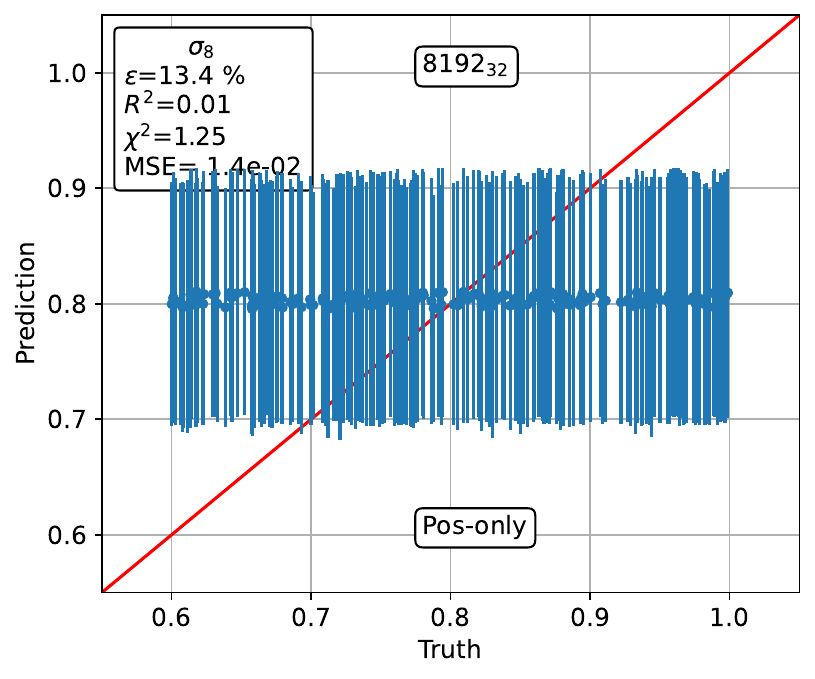}
\includegraphics[width=0.45\linewidth]{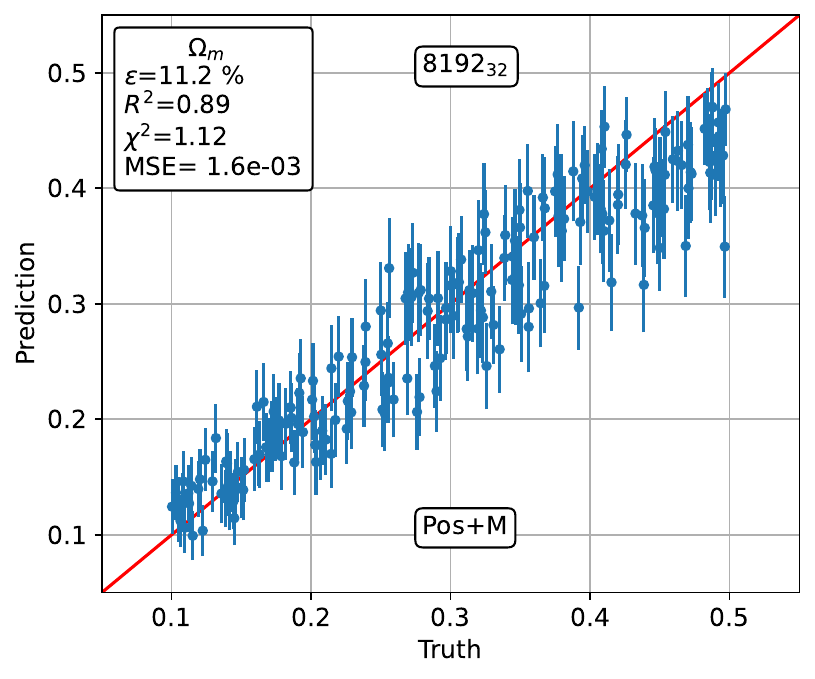}
\includegraphics[width=0.45\linewidth]{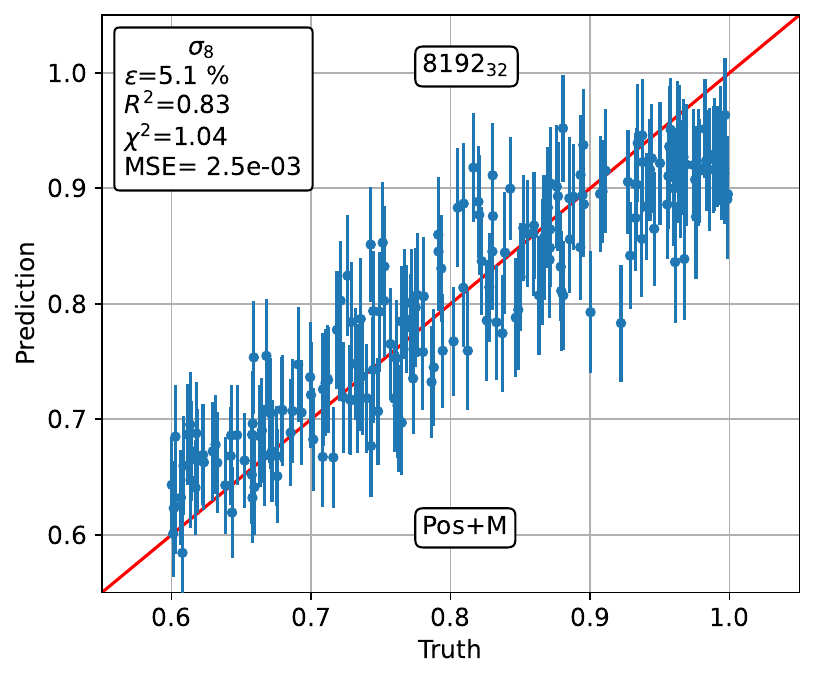}
\includegraphics[width=0.45\linewidth]{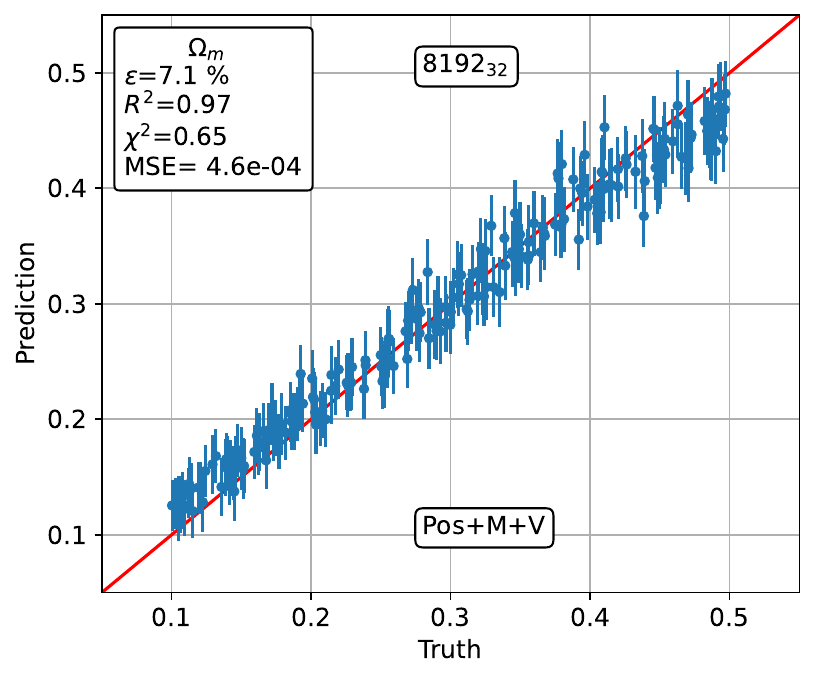}
\includegraphics[width=0.45\linewidth]{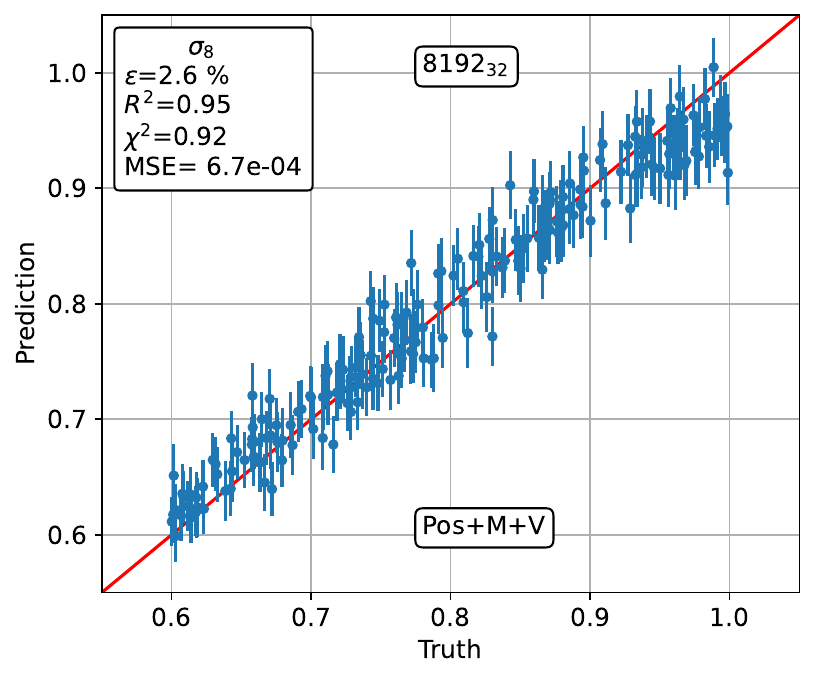}
\caption{Our model is trained to perform likelihood-free inference on both the values of $\Omega_{m}$ (left column) and $\sigma_{8}$ (right column) from point clouds containing: 1) positions (top row), 2) positions and masses (middle row), and  3) positions, masses, and velocity moduli (bottom row). The notation $N_k$ indicates the number of points considered ($N$) and the number of neighbors ($k$). The numbers in the legend indicate the value of different validation metrics.}
\label{fig:Om_sig8}
\end{center}
\end{figure*}

\begin{table*}
\small
%\begin{adjustbox}{max width=\textwidth} % adjust table width
\begin{longtable}{|c|c|c|c|c|c|c|c|c|c|}
\hline
\multirow{2}{*}{Configuration} & \multirow{2}{*}{Features} & \multicolumn{4}{c|}{$\Omega_{m}$} & \multicolumn{4}{c|}{$\sigma_{8}$} \\
\cline{3-10}
& & $\epsilon (\%)$ & $R^2$ & $\chi^2$ & MSE &$\epsilon (\%)$ & $R^2$ & $\chi^2$ & MSE \\
\hline \hline
$1024$ & 2ptCF & 17.4 & 0.75 & 1.05 & $3.4 \times 10^{-3}$ & 11.4 & 0.08 & 1.10 & $1.4 \times 10^{-2}$ \\
\hline
$1024_{32}$ & Pos-only & 21.8 & 0.69 & 0.95 & $4.4 \times 10^{-3}$ & 13.4 & 0.00 & 1.10 & $1.4 \times 10^{-2}$ \\
\hline
$1024_{32}$ & Pos+M & 12.6 & 0.87 & 0.97 & $1.9 \times 10^{-3}$ & 5.2 & 0.81 & 1.06 & $2.6 \times 10^{-3}$ \\
\hline
$1024_{32}$ & Pos+M+V & 10.7 & 0.87 & 0.91 & $1.6 \times 10^{-3}$ & 4.4 & 0.87 & 0.94 & $1.9 \times 10^{-3}$ \\
\hline
$1024$ & Pos+M+V (XgBoost) & 14.8 & 0.68 & - & $3.19 \times 10^{-3}$ & 5.5 & 0.74 & - & $2.84 \times 10^{-3}$ \\
\hline \hline
$4096$ & 2ptCF & 16.2 & 0.81 & 1.04 & $2.6 \times 10^{-3}$ & 8.7 & 0.43 & 1.12 & $7.1 \times 10^{-3}$ \\
\hline
$4096_{32}$ & Pos+M+V & 7.1 & 0.94 & 0.51 & $5.7 \times 10^{-4}$ & 2.7 &  0.95 & 0.52 & $7.2 \times 10^{-4}$ \\
\hline
$4096_{64}$ & Pos+M+V & 8.7 & 0.95 & 0.59 & $7.7 \times 10^{-4}$ & 3.2 & 0.93 & 0.60 & $1.0 \times 10^{-3}$ \\
\hline
$4096$ & Pos+M+V (XgBoost) & 12.6 & 0.78 & - & $2.36 \times 10^{-3}$ & 5.4 & 0.74 & - & $2.8 \times 10^{-3}$ \\
\hline \hline
$8192$ & 2ptCF  & 16.4 & 0.78 & 1.75 & $2.9 \times 10^{-3}$ & 8.5 & 0.46 & 1.43 & $6.7 \times 10^{-3}$ \\
\hline
$8192_{32}$ & Pos-only  & 15.6 & 0.80 & 0.81 & $2.8 \times 10^{-3}$ & 13.4 & 0.01 & 1.25 & $1.4 \times 10^{-2}$ \\
\hline
$8192_{32}$ & Pos+M  & 11.2 & 0.89 & 1.12 & $1.6 \times 10^{-3} $ & 5.1 & 0.83 & 1.04 & $2.5 \times 10^{-3} $ \\
\hline
$8192_{32}$ & Pos+M+V  & 7.1 & 0.97 & 0.65 & $4.6 \times 10^{-4}$ & 2.6 & 0.95 & 0.92 & $6.7 \times 10^{-4}$ \\
\hline
$8192$ & Pos+M+V (XgBoost)  & 11.4 & 0.83 & - & $1.8 \times 10^{-3}$ & 5.2 & 0.76 & - & $2.6 \times 10^{-3}$ \\
\hline
\end{longtable}   
%\end{adjustbox}
\setcounter{table}{0}
\caption{This table summarizes the main results of this work. The first column states the configuration used with the notation $N_k$, where $N$ is the number of halos in the catalog and $k$ is the number of neighbors. The second column shows the features used for the halos, where pos, M, and V stand for positions, masses, and velocity moduli. While the 2ptCF denotes the usage of the 2-point correlation function using just the positions of the dark matter halos, XGBoost indicates the cases where the XGBoost model are used. The other columns show the value of the different validation metrics for $\Omega_{\rm m}$, and $\sigma_8$.}
\label{tab:res_invariant}
\end{table*}
We now present the results we obtained after training the models. Table \ref{tab:res_invariant} summarizes the results for the different configurations. In Fig. \ref{fig:Om_sig8}, we show the constraints we derive on $\Omega_{\rm m}$ and $\sigma_8$ when considering 8,192 halos with 32 neighbors when using different features for the dark matter halos. We now describe the main features we observe:
\begin{itemize}
    \item Pos-only: From table \ref{tab:res_invariant}, we can see that when considering point clouds that only contain the positions of the dark matter halos, the models can infer $\Omega_{\rm m}$ but not $\sigma_8$, even in the case of 8,192 halos. We identify this as one of the main limitations of this model and discuss in the next section (and in Appendix \ref{sec:2pt}) that it may be due to the inability of the network to extract information from small scales. 

    \item Pos+M: When the point clouds contain both halo positions and masses, we find that our models outperform the results of the ones trained with only positions, as expected. Furthermore, in this case, our models can also infer the value of $\sigma_8$. This is likely because the information may come from the halo mass function rather than clustering. Also, in this case, the larger the number of points, the better the performance of the model.

    \item Pos+M+V: We find that the models trained on point clouds that include positions, masses, and velocities perform the best among all the ones studied in this work. As expected, the more points we include, the better the model's performance. In the case of $8132_{32}$, the model is able to constrain $\Omega_{\rm m}$ and $\sigma_8$ with a mean relative error of 7.1\% and 2.6\%, respectively. 
\end{itemize}

%\end{adjustbox}

\subsection{Comparison with 2 point correlation function}

As a benchmark for our analysis, we have quantified how well we can constrain the value of the cosmological parameters if we employ the most standard summary statistics in cosmology: the 2-point correlation function. For this, we have computed the 2-point correlation function of the point clouds (using only halo positions) for different configurations (1,024, 4,096, and 8,192 points) using \texttt{Pylians} \citep{Pylians}. 

We have then trained MLPs to perform parameter inference using these summary statistics. In this case, we use the same loss function as Eq. \ref{Eq:loss}, but the architecture is composed of fully connected layers with LeakyReLU non-linear activation function. The number of layers, neurons, learning rate, weight decay, and dropout are hyperparameters that we optimize with \texttt{OPTUNA}. The results are displayed in Table \ref{tab:res_invariant}.

We find that the MLP trained on the 2-point correlation function outperforms our point cloud model. For 1,024 points, both the MLP and the point cloud model cannot infer $\sigma_8$, while for 8,192 points, the MLP is able to get some loose constraints while the point cloud is unable. We note that this result depends on the minimum scale used in the 2-point correlation function. In the appendix, we show that using the correlation function on scales larger than $10~h^{-1}{\rm Mpc}$ prevents the MLP model from constraining $\sigma_8$. 

Thus, we conclude that when using halo positions alone, our model underperforms the classical 2-pt correlation function. This could be due to multiple reasons, from the lack of training data to our model being suboptimal. We will discuss this in more detail in the next subsection.

\subsection{Comparison with \texttt{XGBoost}}

We also benchmark our study with \texttt{XGBoost}\footnote{\url{https://xgboost.readthedocs.io/en/stable/index.html}} \citep{2016arXiv160302754C}, a simple non-neural random-forest based algorithm. The aim of this exercise is to highlight how a random-forest based regression algorithm with Pos+M+V features performs compared to \texttt{PointMLP-elite}.
A key distinction between \texttt{XGBoost} and \texttt{PointMLP-elite} is that the former relies solely on the Pos+M+V features of the selected halos, without incorporating any information about the underlying topology from the k-nearest neighbors. We also note that XGBoost can only estimate the mean values of the cosmological parameters but not the standard deviation, therefore, we could not provide the $\chi^2$ values in this case.

We have trained the \texttt{XGBoost} model with $\rm n_{estimators}=400$ (size of the forest to be trained),  \texttt{max\_depth} $=10$ (maximum depth of a tree), and  \texttt{tree\_method= `hist'} (tree construction algorithm used in XGBoost) and presented the results in Table \ref{tab:res_invariant}.

We find that the \texttt{XGBoost} model performs significantly worse compared to the \texttt{PointMLP-elite network}. Especially with 8192 halos, the mean squared errors obtained from \texttt{XGBoost} are nearly an order of magnitude higher than those from the \texttt{PointMLP-elite}. Additionally, the mean relative errors from \texttt{XGBoost} are twice as large as those achieved by \texttt{PointMLP-elite}. Therefore, it is evident that \texttt{PointMLP-elite} model has gained significant insight by learning the topology around a halo through feature extraction from its nearest neighbors.

\subsubsection{Component Ablation Study}
In this subsection, we have described in detail the ablation studies demonstrating how the model performs when we changed individual components in \texttt{PointMLP-elite} architecture. We have summarized the result of the ablation study in table-\ref{tab:model_hyperparameter} and described them as follows.

We have mentioned on the last row of Table-\ref{tab:res_invariant}, the values of different validation matrices for both $\Omega_{m}$, $\sigma_8$. To reiterate, we have obtained that result with (a) configuration $8192_{32}$, (b) features Pos+M+V, (c) FPS subsampling, (d) Max-pool aggregation, (e) 4 layers. For the purpose of the ablation study, we kept that configuration as our  benchmark and then varied different individual parts of the \texttt{PointMLP-elite} network as follows

\begin{enumerate}

\item Network depth: We varied the number of stages from 3 to 6. As shown in Table-\ref{tab:model_hyperparameter}, the accuracy of the model declined with 3 stages. The accuracy significantly increased with 4 stages, but then we found that the performance of the model did not improve after 4 stages; only the time required for execution of the codes increased.

\item Max Pooling: we have then tried changing the aggregation method from Max-Pooling to the mean and sum aggregation but found no significant change in the model performance.

\item  Subsampling Method: We have also experimented with the subsampling method. We tried a random subsampling technique rather than the FPS, but the accuracy of the model remains unchanged.
\end{enumerate}
\begin{table*}
\small
%\begin{adjustbox}{max width=\textwidth} % adjust table width
\begin{longtable}{|c|c|c|c|c|c|c|c|c|}
\hline
\multirow{2}{*}{Fixed part} & \multirow{2}{*}{Varying part} & \multirow{2}{*}{Features} & \multicolumn{3}{c|}{$\Omega_{m}$} & \multicolumn{3}{c|}{$\sigma_{8}$} \\
\cline{4-9}
& & & $\epsilon (\%)$ & $R^2$ & MSE  &$\epsilon (\%)$ & $R^2$ & MSE  \\
\hline \hline
 \multirow{4}{*}{Max-pool, FPS} & \multirow{4}{*}{Number of Layers} & 3 layers & 8.3 & 0.89  & $2 \times 10^{-3}$ & 6.2 & 0.75  & $8.9 \times 10^{-3}$\\
\cline{3-9}
 & & 4 layers & 7.1 & 0.97 & $4.6 \times 10^{-4}$ & 5.2 & 0.81 & $2.6 \times 10^{-3}$ \\
\cline{3-9}
 & & 5 layers & 6.9 & 0.95 & $4.9 \times 10^{-4}$ & 5.9 & 0.85  & $2.4 \times 10^{-3}$ \\
\cline{3-9}
 &  & 6 layers & 7.5 & 0.98 & $4.7 \times 10^{-4}$ & 5.5 & 0.81 & $2.3 \times 10^{-3}$ \\
\hline \hline
 \multirow{3}{*}{4 layers, FPS} & \multirow{3}{*}{Aggregation} & Max-pool & 7.1 & 0.97  & $4.6 \times 10^{-4}$ & 5.2 & 0.81 & $2.6 \times 10^{-3}$ \\
\cline{3-9}
  & & Mean & 7.5 & 0.94 & $4.8 \times 10^{-4}$ & 5.0 &  0.79 & $2.5 \times 10^{-3}$ \\
\cline{3-9}
 & & Sum & 7.7 & 0.95 & $5.0 \times 10^{-4}$ & 5.4 & 0.83 & $2.7 \times 10^{-3}$ \\
\hline \hline
 \multirow{2}{*}{4 layers, Max-pool} &\multirow{2}{*}{Subsampling} & FPS  & 7.1 & 0.97 & $4.6 \times 10^{-4}$ & 5.2 & 0.81 & $2.6 \times 10^{-3}$ \\
\cline{3-9}
 & & Random  & 7.2 & 0.95 & $4.2 \times 10^{-3}$ & 5.3 & 0.83 & $2.9 \times 10^{-3}$ \\
\hline

\hline
\end{longtable} 
\setcounter{table}{1}
\vspace{0.5 cm}
\caption{This table summarizes the results obtained by varying different hyperparameters/parts of the network. The first column states the parts of the architecture that were kept fixed, while the second column shows the varying parts of the network. The third column mentions the details of the varying features. For all the scenarios, we use 8192 halos with 32 nearest neighbors, i.e.,$8192_{32}$, with Pos+M+V as the input feature.}
\label{tab:model_hyperparameter}
\end{table*}

\subsection{Comparison with other works}

Now, we will compare our results with those obtained by other groups using different models. We emphasize that a proper comparison can only be made in a few cases where the dataset is the same.

     In \cite{2022OJAp....5E..18M}, the authors used a graph neural network (GNN) along with an Information Maximising Neural Networks on halo catalogs from the \texttt{Quijote} simulations and reported the mean relative errors of $\Omega_{m}$ and $\sigma_{8}$ to be $7.5\%$ and $0.89\%$ when using the positions and masses of $\sim 100$ halos. Our model for $8192_{32}$ with Pos+M performs worse for both parameters. Interestingly, they have also compared their GNN result with the 2ptCF for their halo catalogs and found that the 2ptCF outperforms their GNN models (with a fixed number of halos) for the Pos-only scenario. 
    
      In \cite{2024arXiv240205137H}, a GNN architecture was trained on catalogs containing only the position of $\sim 10,000$ halos from the \texttt{Quijote} simulations to infer cosmological parameters. Although the authors did not report any values of the accuracy metrics, our visual estimate (of Fig. 7 in their paper) suggests that they recover $\Omega_{m}$ with $ \sim 20\%$ which, is slightly poorer compared to our Pos-only study of 8,192 halos. We note that they have been able to infer $\sigma_{8}$ with $\sim 4\%$ accuracy, which our Pos-only study fails to do. 

     In \cite{Carolina_diffusion_generative_modelling}, the authors used a diffusion-based generative model for generating point clouds that resemble the spatial distribution of dark matter halos from an N-body simulation. These authors used the 5,000 most massive halos of the latin-hypercube \texttt{Quijote} simulations to train their model. Once trained, their model can also be used to infer the likelihood of the data. They quantified the performance of their model in two scenarios: (i) Pos-only and (ii) Pos+M+Vel (Vel here being the 3D velocity). In the case of Pos-only, they achieved a mean relative error of $\sim5\%$ and $\sim3\%$ on $\Omega_{\rm m}$ and $\sigma_8$, respectively. However, the learned likelihood from their model is not well calibrated for $\Omega_{\rm m}$, so comparison with that cosmological parameter should be taken with a grain of salt.
    
     In \cite{2022ApJ...937..115V}, the authors use a GNN-based approach to infer the value of cosmological parameters $\Omega_{m}$ and $\sigma_{8}$ based on the galaxy catalogs from the CAMELS simulations \citep{CAMELS_presentation}. Further, they study three scenarios based on the input feature used - (i) only position of the galaxy, (ii) galaxy positions + stellar masses + stellar radii + stellar metallicities, and (iii) galaxy positions + stellar masses + stellar radii + stellar metallicities + $\rm V_{max}$. In this case, the authors were able to infer the value of $\Omega_{\rm m}$ with different degrees of confidence. We note that the catalogs, volume, and nature of the simulations are very different from the ones used in this work, so a comparison is not possible.
    
     In \cite{shao_et_al}, the authors used a similar GNN-based model to infer $\Omega_{m}$ and $\sigma_8$ separately with $\leq 5000 $ halos from halo-catalogs obtained from Gadget N-body simulations in a periodic volume of $(25 \rm h^{-1} Mpc)^3$. Two cases from their study that are relevant for comparison with the current works are: (i) Pos-only for inferring $\Omega_{m}$ and $\sigma_8$ (ii) Pos+M for inferring $\sigma_{8}$. For (i) Pos-only, they obtained a mean relative error of $10 \%$ and $11.8 \%$ for $\Omega_{m}$ and $\sigma_8$, respectively. In our case for Pos-only and using 8192 halos, we obtained a mean relative error of $15.6\%$ for $\Omega_{m}$. For Pos+M, they obtained a mean relative error of $5.9\%$ for $\sigma_{8}$, whereas we obtained a mean relative error of $7.1\%$ for $\sigma_{8}$ with 8192 halos. Here, also note that their networks infer only $\sigma_{8}$, unlike ours, which infers both $\Omega_{m}$ and $\sigma_8$ at the same time. We would like to remind the readers that they used a box size of $(25 \rm h^{-1} Mpc)^3$, but in our case, the box size is $(1000 \rm h^{-1} Mpc)^3$, so a proper comparison is not possible and this should be seen as a qualitative statement. 
    
     In \cite{2022arXiv221112346A}, the authors employ a \texttt{PointNeXt} network using halo catalogs from COSMO-GRIDV1 simulations \citep{CosmoGrid} with (i) Pos-only and (ii) Pos+M as the input features to infer the values of $\Omega_{m}$ and $\sigma_{8}$. Their MSE for $\Omega_{m}$ with 8,000 halos using Pos-only is $3.6 \times 10^{-3}$ which is slightly poorer compared to our $8192_{32}$ Pos-only scenario ($2.8 \times 10^{-3}$). Further, with the same halo number, their Pos+M model yields an MSE of $1.6 \times 10^{-3}$, which is slightly better than the MSE obtained from our Pos+M scenario $2.5 \times 10^{-3}$. Note that unlike us, their Pos-only case outperforms the 2ptCF result only in the case of $\sigma_8$ (as seen from Fig. 2b of their paper). We note that also, in this case, a proper comparison is not possible given the different setups (lightcone versus periodic boxes) and volumes/resolutions of the catalogs.

%%%%%%%%%%%%%%%%%%%%%%%%%%%%%%%%%%%%%%%%
%%%%%%%%%%%%%%%%%%%%%%%%%%%%%%%%%%%%%%%%
\section{Model limitations and improvements}
\label{sec:discussion}

We now discuss some of the limitations of our model and potential directions to overcome them.

 Perhaps the most important limitation of the model studied here is its inability to infer $\sigma_8$ from the points clouds when using only the positions of the halos. Other works using different architectures have been shown to provide relatively good constraints on $\sigma_8$ using halos from the \texttt{Quijote} simulations \citep[see][]{2024arXiv240205137H, 2023arXiv231117141C}. This may be because the model cannot properly extract information from small scales, where the constraints from $\sigma_8$ most likely arise (see Appendix \ref{sec:2pt}). Below, we discuss a few potential architecture changes that could help improve this behavior. 

 Another limitation is that standard summary statistics like the 2-point correlation function seem to contain more information than the one our model can extract when using point clouds with only the positions of the halos. This may be related to the previous point (i.e., the inability of our model to extract small-scale information). On the other hand, the same behavior has been seen in \cite{2021JCAP...11..049M} using the same dataset but with more expressive GNN architecture. Thus, perhaps the limitation in this case is the limited training dataset (only 1,600 point clouds). In the future, it would be interesting to train our models with larger datasets to see if the constraints on the parameters improve. Another way to overcome this problem would be to concatenate the 2-point correlation function measured on the halo catalogs with the latent space before passing it into the inference model \cite[see e.g.][]{Michelle_2020}.

 Another potential limitation is that in the current model, the input point clouds must have the same fixed number of points. While this may not be a problem if one has enough halos and performs a cut in number density, there may be applications where a selection function may yield different numbers of halos in different catalogs (e.g., a cut in halo mass). Modifications to our model can enable working with point clouds of different sizes. For instance, in the first stage, one may sample a fixed number of points using the farthest point sampling algorithm rather than just taking half of the points to the input layer.

 As a proof-of-concept, we have carried out this study using dark matter halos in real-space together with their masses and velocity moduli. To get closer to the data, we will need to work with galaxies in redshift-space, or galaxies in real-space but with radial peculiar velocities \citep[see e.g.][]{Natali_2023}. Besides, we need to place the halos/galaxies in a lightcone rather than in periodic cubic boxes. It will be interesting to repeat the work done in this paper with these more realistic setups.

 Another potential improvement would be to consider different neighborhood definitions. For instance, in this work, we use the k nearest neighbors of points in each stage following \cite{2022arXiv220207123M}. However, \cite{pablo_villanueva_domingo_2022_6485804} showed that using all points within a given distance yields better results than using the k closest neighbors when training GNNs. Thus, it would be interesting to try that neighborhood definition and investigate the results.

\section{Conclusions}
\label{sec:conclusions}

In this paper, we have trained neural networks to perform likelihood-free inferences on the value of the cosmological parameter from halo catalogs, which are represented as point clouds. This task has been carried out before using different approaches like graph neural networks (GNNs) \citep[see, e.g.][]{pablo_villanueva_domingo_2022_6485804, Helen_2022, 2021JCAP...11..049M, 2022arXiv221112346A, Natali_2023, 2024arXiv240205137H, 2023arXiv231117141C}. The main difference in our work is that we use a recently proposed architecture, pointMLP \citep{2022arXiv220207123M}, that can work directly with point clouds and is able to extract information hierarchically for large datasets. As with GNNs, the advantage of using these models is that they can extract information at the field level without relying on summary statistics. Besides, they do not impose a minimum scale where the field can be sampled, as happens when the point clouds are transformed into regular grids. Furthermore, they can automatically deal with complex geometric patterns, such as selection effects and masks, that are more difficult to treat with standard approaches.

The model takes a point cloud containing $N$ points as input, reducing the number of those hierarchically while increasing the dimensionality of their feature vector. Finally, an MLP is used to infer the parameters from the points in the last hierarchical layer. The features of the points in the hierarchy are updated from their neighbors using  MLPs and max pooling aggregation. This model performs similar operations as GNNs but without requiring an input graph. 

We find that our models perform well and can accurately infer the value of the cosmological parameters under different combinations of the input features (e.g., positions only, positions plus masses...etc). A comparison with other works in the literature is difficult due to intrinsic differences in the training data and methodology. Perhaps the closest analyses are the ones from \cite{2024arXiv240205137H}, \cite{2022OJAp....5E..18M}, and \cite{Carolina_diffusion_generative_modelling}. Our model yields better and worse results for $\Omega_{\rm m}$ depending on setup and parameter. However, all these works perform better when inferring $\sigma_8$ than our model. 

The model explored in this work presents several advantages with respect to previous methods (e.g., message-passing GNNs). First, its simplicity and lack of sophisticated geometric kernels make the model fast to evaluate. Second, the model can work with large point clouds. In this work, we have explored models that contain up to 8,192 points, but this number can easily be made much larger if needed (e.g., decreasing the number of points to sample during the stages).
\footnote{We would like to \textbf{emphasize} that we dealt with 80,000 halos (when the periodic boundary condition is ignored) in this work which is not even feasible for GNN. Thus, for lightcone datasets, our model can easily deal with hundreds of thousands of points if several GPUs are used.} This low computation time for PointMLP is extremely important in the context of future sky surveys. In the upcoming sky surveys, the number of the galaxy will be very large, and dealing with a large number of galaxies using GNN is not possible even with state-of-the-art computers. Therefore, a model like PointMLP with a computational time less than that of \textbf{a GNN} will be very important for future galaxy surveys. Third, its simplicity makes it very easy to adapt to different tasks. Fourth, this model works with the natural representation of the data (a point cloud), while GNNs need a graph (it is unclear how to construct the optimal graph from a point cloud) to operate. We note that deep sets are deep learning architectures designed to work with data that can be represented as sets, but in general, they do not exploit the spatial information \cite[see e.g.][for an example of using deep sets to infer cosmology from void catalogs]{Bonny_2023}.

On the other hand, we find that in the simplest case of a point cloud containing only the positions of the halos, the classical 2-point correlation function yields better results. While this can be due to multiple reasons, such as lack of training data and low number of halos (high shot-noise), it can also be due to the architecture not being able to exploit the clustering information fully. We note that this result has also been obtained using a GNN architecture in \cite{2022OJAp....5E..18M}, perhaps pointing towards the low-data directions rather than the expressivity of the model. We have identified some areas that could enhance the expressivity of our model and leave this for future work.

Finally, we note that while in this work we have worked with halo catalogs from N-body simulations, the method can be applied to other point cloud data, like galaxy catalogs from hydrodynamic simulations or semi-analytic models \citep{2022arXiv220101300V, 2022arXiv220402408P} or void catalogs \citep{Gigantes}. The main reason for choosing the halo catalogs rather than the galaxy catalogs is the unavailability of training data sets with volumes comparable to the upcoming sky surveys. The state-of-the-art Hydrodynamical data suit available to us is CAMELS, which covers a volume of $(25 h^{-1} Mpc)^3$, much smaller than the volume of future sky surveys. As this work is a precursor to those sky-surveys data, we chose Quijote simulation with a box size of 1 $h^{-1}$ Gpc, comparable to the upcoming sky surveys. \\

\section*{Acknowledgments}
We thank Teresa Huang, Natali de Santi, and Matthew Ho for their comments and discussion on this work. The calculations performed in this work have been carried out in both the IUCAA cluster and Rusty cluster at the Flatiron Institute. The work of FVN is supported by the Simons Foundation.

\appendix

\section{2 point correlation function}
\label{sec:2pt}

In this appendix, we study the dependence on the constraints we obtain using the 2-point correlation function as a function of the minimum scale used. We consider two different scenarios: 1) we use the correlation function on all available scales, and 2) we only use the correlation function on scales larger than $10~h^{-1}{\rm Mpc}$. For each of these cases, we train MLPs from scratch and perform hyperparameter tunning as described in Section \ref{sec:methods}.

\begin{figure*}[th!]
\begin{center}
\includegraphics[width=0.45\linewidth]{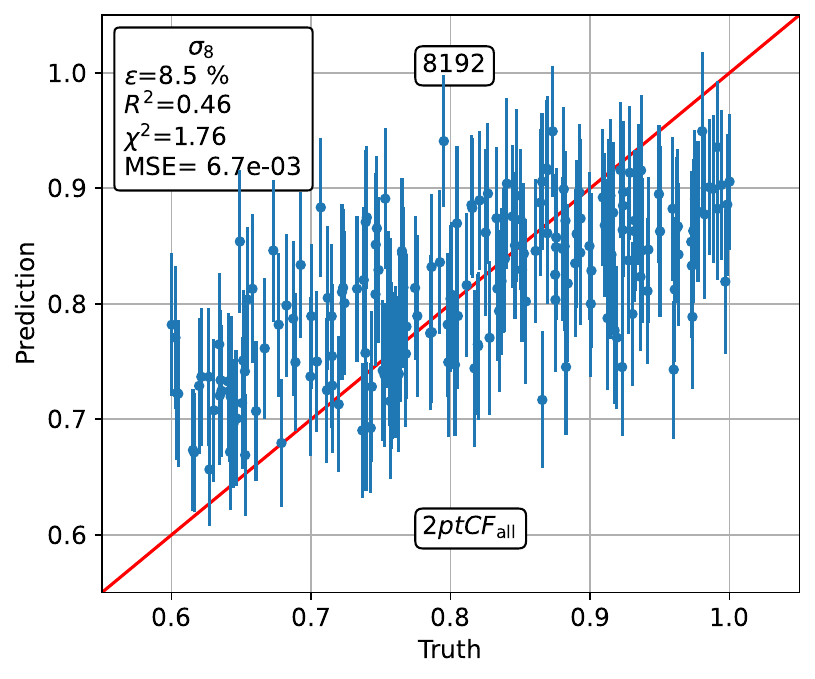}
\includegraphics[width=0.45\linewidth]{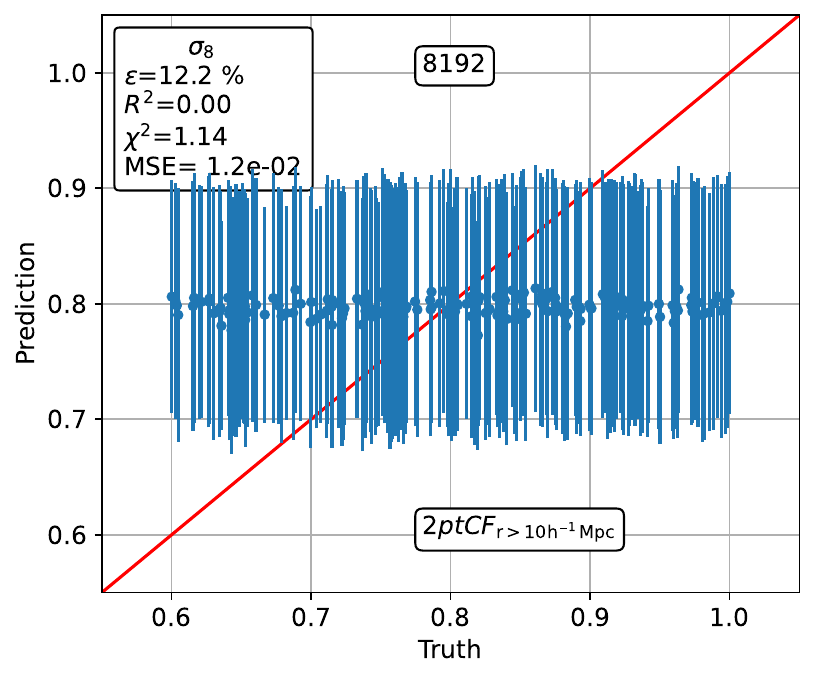}
\includegraphics[width=0.45\linewidth]{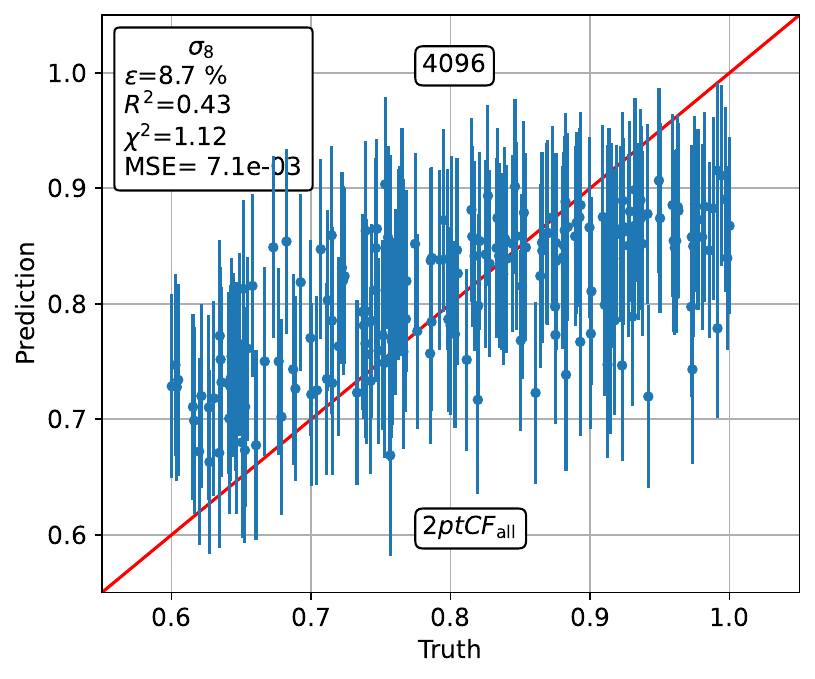}
\includegraphics[width=0.45\linewidth]{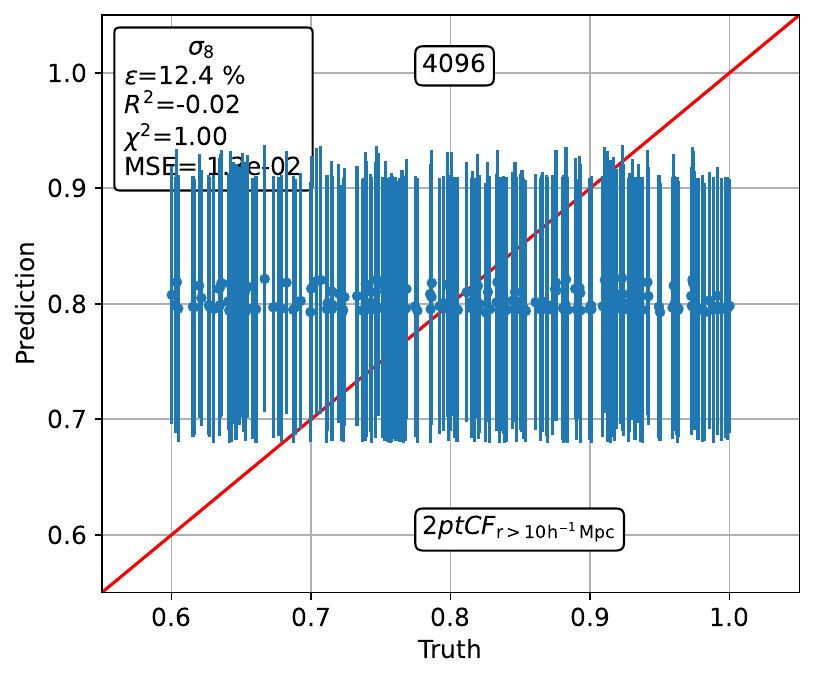}
\includegraphics[width=0.45\linewidth]{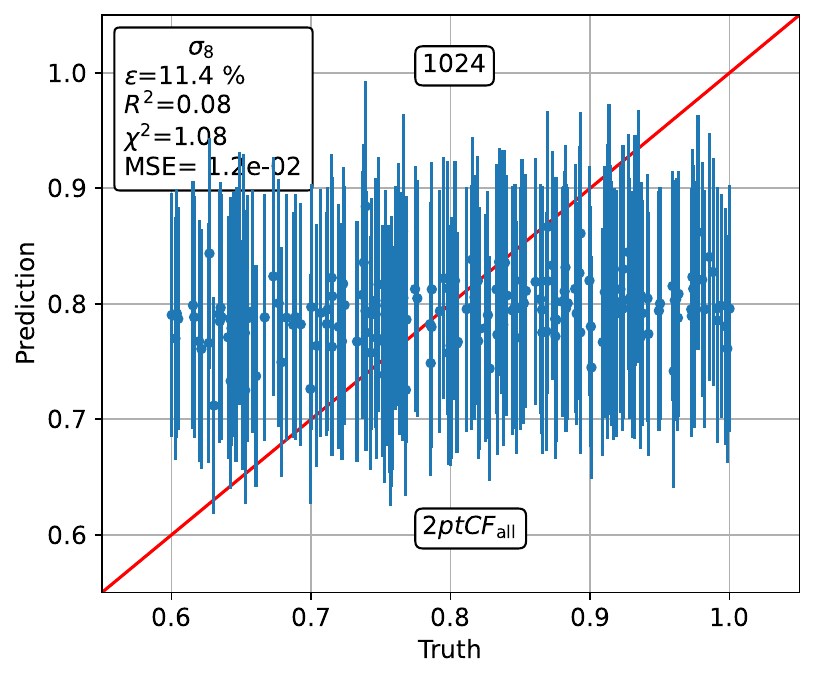}
\includegraphics[width=0.45\linewidth]{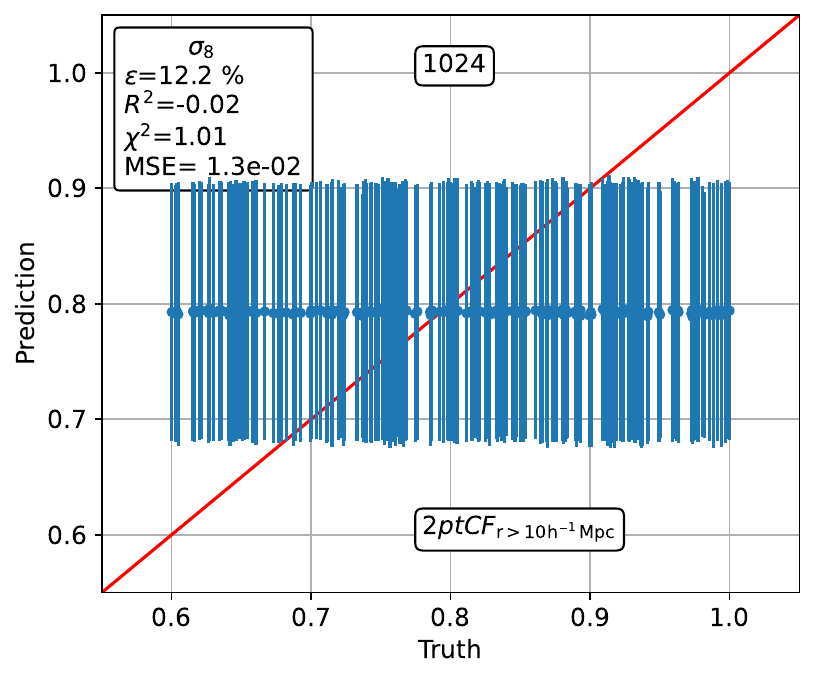}

\caption{From point clouds that only contain the positions of the dark matter halos, we have measured their 2-point correlation functions. We have then trained MLPs to infer the value of $\Omega_{\rm m}$ and $\sigma_8$ from them. This plot shows the results of this analysis when using 8,192 (top), 4,096 (middle), and 1,024 (bottom) halos employing all scales from the correlation function (left column) and only scales larger than $10~h^{-1}{\rm Mpc}$ (right column). As can be seen, to infer $\sigma_8$ we need to use scales larger than $10~h^{-1}{\rm Mpc}$.}
\label{fig:sig8_2ptCF}
\end{center}
\end{figure*}

We show the results of this analysis in Fig. \ref{fig:sig8_2ptCF}. We can reach two important conclusions. First, the larger the number of points, the more accurate the network becomes. This is clearly expected, given the fact that the shot noise will be lower. Second, we find that if we do not use scales smaller than 10 $h^{-1}{\rm Mpc}$, the network cannot infer the value of $\sigma_8$. This is perhaps expected as $\sigma_8$ measures the variance of the linear perturbations in the underlying matter field on scales of $8~h^{-1}{\rm Mpc}$.

This behavior may shed some light on why our point could model does not yield good results for $\sigma_8$. Perhaps the above finding points towards the point cloud model not being able to exploit the small-scale information optimally. We speculate that this may be due to the hierarchical nature of the model, the pooling layer, or the subsampling method. We note that other works using the same data managed to get good constraints on $\sigma_8$ using a similar number of halos to the ones considered in this paper \citep[see][]{2024arXiv240205137H, 2023arXiv231117141C}.

\bibliography{main}{}
\bibliographystyle{aasjournal}

\end{document}